\documentclass[conference]{IEEEtran}

\usepackage{cite}
\usepackage{graphicx}
\usepackage{url}
\usepackage{xcolor}
\usepackage{float}

\usepackage{tikz}
\usetikzlibrary{shapes, arrows.meta, positioning, decorations.pathmorphing}

\newcommand{\remove}[1]{}

\begin{document}

\title{Is Crunching Public Data the Right Approach to Detect BGP Hijacks?}

\author{
\IEEEauthorblockN{Alessandro Giaconia\IEEEauthorrefmark{1},
Muoi Tran\IEEEauthorrefmark{2}, 
Laurent Vanbever\IEEEauthorrefmark{1},
Stefano Vissicchio\IEEEauthorrefmark{3}
}

\IEEEauthorblockA{\IEEEauthorrefmark{1}ETH Zürich, \emph{\{agiaconia, lvanbever\}@ethz.ch}}
\IEEEauthorblockA{\IEEEauthorrefmark{2}Chalmers University of Technology and University of Gothenburg, \emph{muoi@chalmers.se}}
\IEEEauthorblockA{\IEEEauthorrefmark{3}University College London, \emph{s.vissicchio@ucl.ac.uk}}
}

\IEEEoverridecommandlockouts
\maketitle

\begin{abstract}
  The Border Gateway Protocol (BGP) remains a fragile pillar of Internet routing.
  BGP hijacks still occurr daily.
  While full deployment of Route Origin Validation (ROV) is ongoing, attackers have already adapted, launching post-ROV attacks such as forged-origin hijacks.
  To detect these, recent approaches like DFOH~\cite{dfoh} and BEAM~\cite{beam} apply machine learning (ML) to analyze data from globally distributed BGP monitors, assuming anomalies will stand out against historical patterns.
  However, this assumption overlooks a key threat: BGP monitors themselves can be misled by adversaries injecting bogus routes.

  This paper shows that state-of-the-art hijack detection systems like DFOH and BEAM are vulnerable to data poisoning.
  Using large-scale BGP simulations, we show that attackers can evade detection with just a handful of crafted announcements beyond the actual hijack.
  These announcements are indeed sufficient to corrupt the knowledge base used by ML-based defenses and distort the metrics they rely on.
  Our results highlight a worrying weakness of relying solely on public BGP data.

    
\end{abstract}

\IEEEpeerreviewmaketitle

\section{Introduction}

The Border Gateway Protocol (BGP) serves as the backbone of the Internet, directing traffic between the vast network of Autonomous Systems (ASes) that comprise the global infrastructure. 
However, BGP was designed in an era prioritizing connectivity over security, leaving it without inherent mechanisms to verify the authenticity of routing information~\cite{bgpsecuritysurvey}. 
This fundamental vulnerability makes it susceptible to malicious manipulation, most notably BGP hijacking, where an attacker illegitimately claims ownership of IP prefixes, and route leaks, where routing announcements are propagated beyond their intended scope. 
These incidents can redirect significant amounts of internet traffic, enable espionage, cause widespread service outages, and undermine the stability of the Internet \cite{rostelecom, ytpakistan, klayswap, myetherwallet}.

To combat these threats, the Resource Public Key Infrastructure (RPKI) was developed \cite{rpki-rfc}. 
RPKI enables Route Origin Validation (ROV), a way to cryptographically verify that an AS is authorized to originate routes for specific IP prefixes, which mitigates traditional prefix hijacks. 
However, ROV deployment is far from universal~\cite{rodday2023resource, rpki}, leaving large parts of the Internet unprotected. 
Furthermore, RPKI is still vulnerable to some kinds of BGP hijacks, such as forged-origin attacks~\cite{artemis}. 
In these attacks, an adversary announces a prefix belonging to a victim but prepends the victim's legitimate AS to the path. 
This makes the announcement appear valid to RPKI checks while still allowing the attacker to attract traffic.
Another cryptographic solution, BGPSec \cite{bgpsec-rfc}, aims to secure the entire path of BGP announcements, offering stronger protection against path manipulation attacks. 
However, similar to RPKI, its deployment has been extremely slow, limiting its practical effectiveness in the current Internet landscape.

Given the limitations of RPKI and the slow adoption of BGPSec, the BGP security community has increasingly turned towards BGP monitoring. Mainstream approaches leverage data from public collectors like RouteViews \cite{routeviews} and RIPE RIS \cite{ripe-ris}, which aggregate BGP updates from hundreds of vantage points worldwide. 
Recently, Machine Learning (ML) and data-driven techniques have been integrated into this design. 
By processing historical and real-time BGP data, ML-based systems analyze multiple features (e.g., AS-path characteristics, topological changes) across the collected routes to identify anomalous ones, such as BGP hijacks and route leaks. 
Prominent examples of data-driven BGP monitors include ARTEMIS \cite{artemis}, DFOH \cite{dfoh}, and BEAM \cite{beam}, as well as commercial offerings like Cisco's ThousandEyes~\cite{ciscoThousandEyes}.

The effectiveness of ML-based systems however depend on the quality and integrity of their input data, and previous work has already shown the risks of inputs deliberately altered to cause misclassification~\cite{adversarialexamples, Biggio_2018}.
Can BGP hijackers corrupt the knowledge base of ML-based defenses by injecting (a few) BGP routes beyond the actual attack?

This paper demonstrates that state-of-the-art hijack detectors are susceptible to adversarial data manipulation.
Their effectiveness can be significantly compromised if their input BGP data is deliberately manipulated to \emph{not} reflect the ground truth.
Attackers aware of these detectors can indeed craft BGP routes specifically designed to deceive the ML-based defenses.

We show how attackers can poison the knowledge bases used by these defenses or pollute the metrics they rely on, allowing malicious hijacks (including RPKI-compliant forged-origin attacks) to evade detection.
Our simulations show that attackers can easily and stealthily poison monitor data and make monitor-based systems like DFOH~\cite{dfoh} and BEAM~\cite{beam} ineffective: attackers need only to inject a few false BGP announcements in addition to the hijack, and they can compute such announcements quickly on inexpensive hardware. 
Relying on monitor-based ML defenses may thus lead to a false sense of security.
More in general, our findings underscore the need for more robust defenses to protect the Internet's routing ecosystem from motivated adversaries.

\clearpage

Our contributions can be summarized as follows.
\begin{itemize}
    \item We identify several limitations of using ML on data from public BGP monitors to detect BGP hijacks. 
    \item We demonstrate how an adversary can exploit such limitations to evade state-of-the-art hijack detectors, namely DFOH and BEAM, cheaply and stealthily.
    \item We discuss short-term and long-term solutions to this new class of attacks, some of which are readily deployable.
\end{itemize}

\section{Background}
\label{sec:background}

\subsection{Prefix Hijacking}
\label{subsec:hijack}
Prefix hijacking is a malicious activity where an attacker illegitimately claims ownership of IP address blocks (prefixes) they do not control. This is possible because the BGP protocol, which manages routing between ASes, was designed with an inherent trust model, lacking built-in mechanisms to verify the authenticity of route announcements. When an attacker successfully hijacks a prefix, they can redirect internet traffic intended for the legitimate owner, leading to severe consequences such as service outages, espionage, and financial losses, as seen in attacks targeting cryptocurrency platforms like KlaySwap~\cite{klayswap} and MyEtherWallet~\cite{myetherwallet}.

\begin{figure}[t!]
    \includegraphics[width=1\columnwidth]{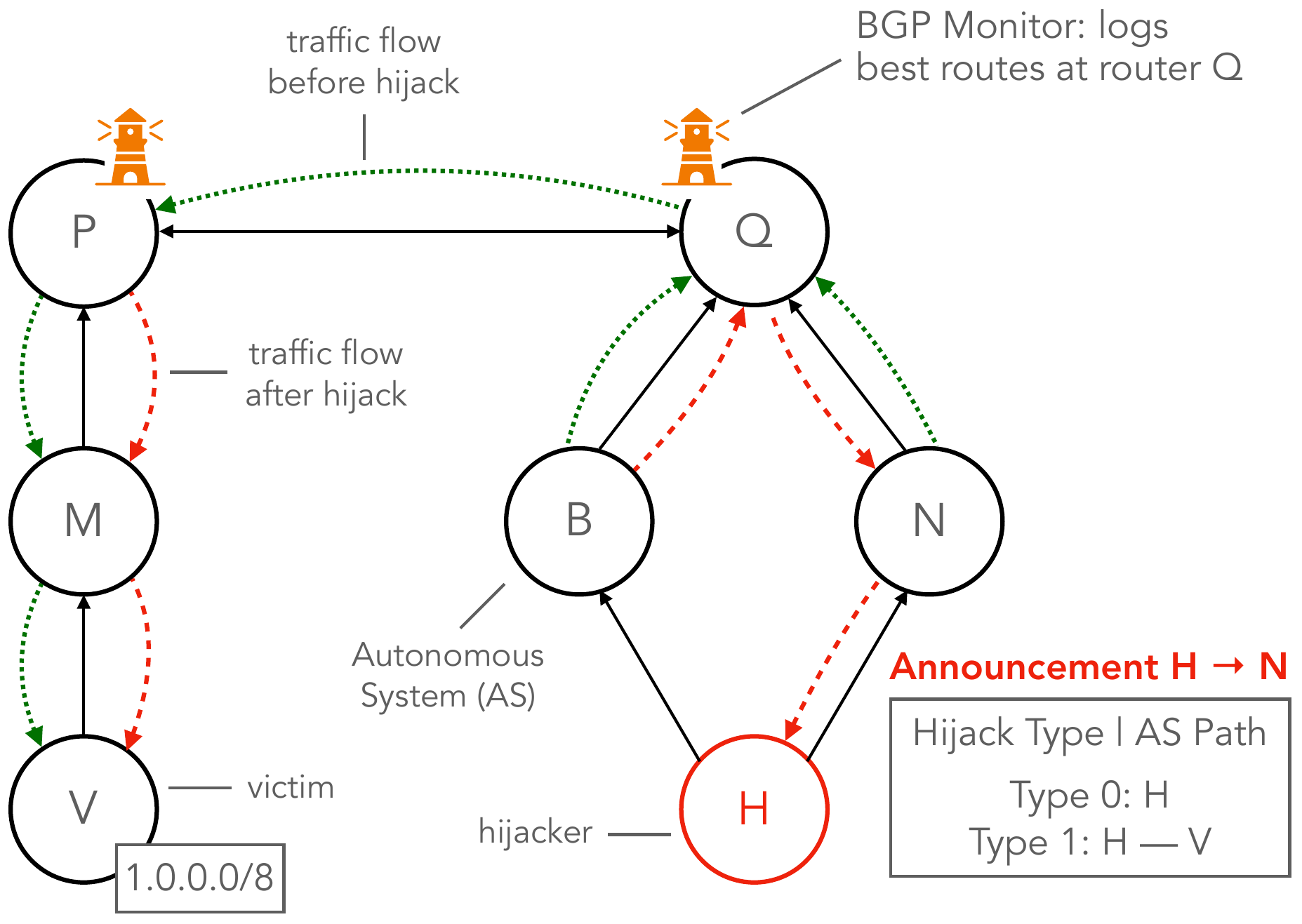}
    \caption{Example of hijack for prefix 1.0.0.0/8 owned by AS \textit{V}. Single-headed solid arrows indicate the direction of the money in customer-provider links; double-headed arrows represent charge-free links.}
    \label{fig:example-hijack}
\end{figure}

Hijacks can be categorized based on their complexity. Following the taxonomy proposed in~\cite{artemis}, a \textit{Type-0 hijack} involves an attacker announcing a prefix belonging to another AS without altering the origin AS in the announcement. This is the simplest form of hijack. A \textit{Type-1 hijack}, or forged-origin hijack, occurs when the attacker announces the victim's prefix but lists the victim's AS as the origin. While the Resource Public Key Infrastructure (RPKI)~\cite{rpki-rfc}, which cryptographically links prefixes to their legitimate origin ASes, can effectively prevent Type-0 hijacks when deployed, it is less effective against Type-1 hijacks if the attacker manipulates the AS path in specific ways (e.g., prepending the legitimate origin). Figure~\ref{fig:example-hijack} illustrates both Type-0 and Type-1 hijack scenarios.

\subsection{BGP monitors}
\label{subsec:bgpmonitors}
BGP monitors are crucial infrastructure components for observing the global state of internet routing. Prominent examples include the RouteViews project~\cite{routeviews} and the RIPE Routing Information Service (RIS)~\cite{ripe-ris}. These projects operate globally distributed collectors that establish BGP peering sessions with numerous ASes worldwide, acting as vantage points. By passively listening to the BGP updates (including announcements and withdrawals) exchanged in these sessions, monitors collect vast amounts of routing data. This data typically includes periodic snapshots of the Routing Information Base (RIB) from their peers and continuous streams of UPDATE messages reflecting real-time changes. The data is typically stored and distributed in the Multi-threaded Routing Toolkit (MRT) format~\cite{mrt-format} and access is provided through near real-time streams, often using the BGP Monitoring Protocol (BMP)~\cite{bmp-rfc}, and periodic archival dumps (e.g., RIB snapshots every few hours and update logs every few minutes). While invaluable, this data provides an incomplete view of the internet's routing dynamics, as monitors only capture routes propagated towards their specific vantage points. Despite this limitation, in the current data-driven era, BGP monitors are indispensable for security analysis, performance monitoring, and research, offering the most comprehensive publicly available source of information for understanding BGP behavior and detecting anomalies like hijacks.

\subsection{Data-driven Hijack Detection}
\label{subsec:hijack-monitor}

Given the limitations of inherent BGP security and the incomplete deployment of RPKI, data-driven hijack detection systems have emerged as a critical layer of defense. These systems analyze BGP data, primarily sourced from public monitors, to identify suspicious routing events.

\begin{figure}[t!]
    \includegraphics[width=0.85\columnwidth]{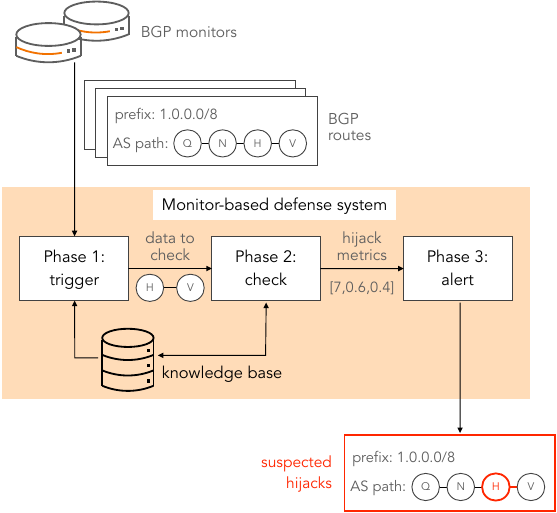}
    \caption{General architecture of a monitor-based hijack detection system.}
    \label{fig:defense-model}
\end{figure}

More sophisticated systems employ machine learning (ML) and anomaly detection techniques to uncover complex attacks, including forged-origin hijacks (Type-1) and route leaks, which might evade simpler checks. These systems analyze features extracted from BGP updates, such as AS-path characteristics, prefix propagation patterns, and consistency with historical data, to build models of normal routing behavior and flag significant deviations \cite{shapiradeeplearninghijackdet, dfoh, beam, artemis, weaklysupervised}. A general overview of such a system is depicted in Figure~\ref{fig:defense-model}. Typically, these systems operate in three phases: first, they are \textit{triggered} by new BGP updates observed by monitors; second, they \textit{check} the legitimacy of the update against a knowledge base of historical data or learned models; finally, if the calculated metrics exceed certain thresholds, they raise an \textit{alert}. For instance, DFOH is triggered by new AS links, checks them by computing features (e.g., topological, PeeringDB-based) against its historical graph, and alerts based on a Random Forest classifier's output. While their specific architectures and algorithms vary, a common characteristic is their reliance on historical BGP announcements collected by monitors to establish a baseline or knowledge base against which new routes are evaluated. This reliance on potentially manipulable public data is a key focus of this paper.

\section{Overview}
\label{sec:overview}

\subsection{Threat Model}
\label{subsec:threat-model}

We consider an adversary who controls an AS and can therefore inject malicious BGP announcements into the global routing system with the intent of disrupting Internet traffic, specifically through prefix hijacking.
We assume that some ASes use RPKI and apply ROV.
This implies that all the BGP announcements injected by the attackers must be RPKI-compliant.
Consistently, the attacker’s primary objective is to execute a Type-1 hijack, which cannot be mitigated by RPKI.

In addition to announcing BGP routes from their AS, the adversary can leverage readily available commercial transit services, often offered by hosting providers~\cite{servperso,ifogch}.
These services allow users to acquire BGP sessions easily and cheaply, frequently through Virtual Machines (VMs).
The attacker can use such additional BGP sessions to announce poisoning BGP routes as detailed in Section~\ref{subsec:poisoning-monitors}.
Our attacks, however, do not directly implicate these additional transit services in the hijacks themselves, thus avoiding their specific abuse detection mechanisms (e.g., prefix filtering).

\subsection{Fundamental Limitations of BGP Monitors}
\label{subsec:fundamental-limitation}

Monitor-based defenses suffer from the following limitations inherent to the collection and analysis of public BGP data.

\begin{itemize}

\item \textit{Untruthful BGP routes cannot be distinguished from truthful ones.}
Data in BGP routes is not validated -- except for the ownership of the originated prefix, if we assume RPKI.
Hence, for any newly observed BGP route, at least two plausible scenarios always exist: (i) the route includes some fake data, or (ii) the route reflects an actual routing change resulting from new business agreements, updated BGP configurations, or topological changes (e.g., failures). These two scenarios are fundamentally indistinguishable from the perspective of external observers. As a result, monitor-based defenses must rely on intrinsically inaccurate heuristics based on factors like AS geography or inferred business relationships to evaluate the truthfulness of data (e.g., the AS path) in BGP routes.

\item \textit{Hijackers can manipulate routes collected by BGP monitors.}
Any AS can announce BGP routes with some fake data: this is demonstrated and leveraged in BGP hijacks themselves.
Since untruthful data in BGP routes cannot be detected (see previous point), BGP monitors log all the BGP routes they receive, including the ones including fake data.
This means that monitor-based defenses must rely on a knowledge base that may not reflect the ground truth of the Internet ecosystem.

\item \textit{It is impossible to identify invalid or expired BGP data.}
When a new legitimate BGP route is observed, there is fundamentally no way to determine how long it should be remembered as valid. By design, BGP does not signal when paths or links do not exist anymore. So, at any time, previously announced paths may still be valid and just not being currently used, or they might have actually expired. Also, BGP paths vary greatly in their stability over time: some persist for months while others are very short-lived. Monitor-based defense systems must thus make arbitrary decisions about data retention.

\end{itemize}

Collectively, these limitations create a fundamental vulnerability: monitor-based defenses are forced to operate on data that may be manipulated by potential attackers, with no reliable way to detect when this occurs or to decide that outdated data has to be discarded. This vulnerability creates opportunities for attackers to evade monitor-based defenses, as we elaborate hereafter.

\subsection{Poisoning BGP Monitors}
\label{subsec:poisoning-monitors}

Consider an attacker aiming to poison the knowledge base of monitor-based defenses, in preparation for a hijack that would otherwise be detected by such defenses.

The first question is whether the attacker can ensure that their BGP routes actually reach the collectors used by these defenses, and if so, how.
The answer to this question is not straightforward, given that ASes not controlled by the attacker can filter routes based on RPKI records and custom route filters they configure.
For example, external ASes can filter received routes based on the announced AS path or IP prefix.

We now describe a general methodology that attackers can use to announce BGP routes that include a fake link in the AS path and avoid common route filters deployed in Internet ASes.

In their poisoning of BGP routes, the attacker can use a sub-prefix $p'$ of a prefix legitimately owned by the AS of the attacker.
To create an AS path with a fake link, the attacker can then specify that the AS originating the route for $p'$ is an external, strategically chosen AS B.
Doing so avoids prefix filters that providers of the attacker's AS may have: the attacker's AS could have actually sold $p'$ to B, so the route must not be filtered.

To avoid RPKI-based filters, the attacker ensures that $p'$ does not have a Route Origin Authorization (ROA), or creates a ROA that includes the fake origin AS B. For ROV-enabled ASes, the BGP announcement would have a 'NotFound' RPKI validation state. Since RPKI deployment is still far from universal and many prefixes lack ROAs~\cite{rpki}, even ROV-enabled ASes are expected to not drop routes with 'NotFound' status, as also recommended by RFC 7115~\cite{rfc7115}.
This should ensure the propagation of the poisoning announcement Internet-wide.


\begin{figure}
    \centering
    \includegraphics[width=1\columnwidth]{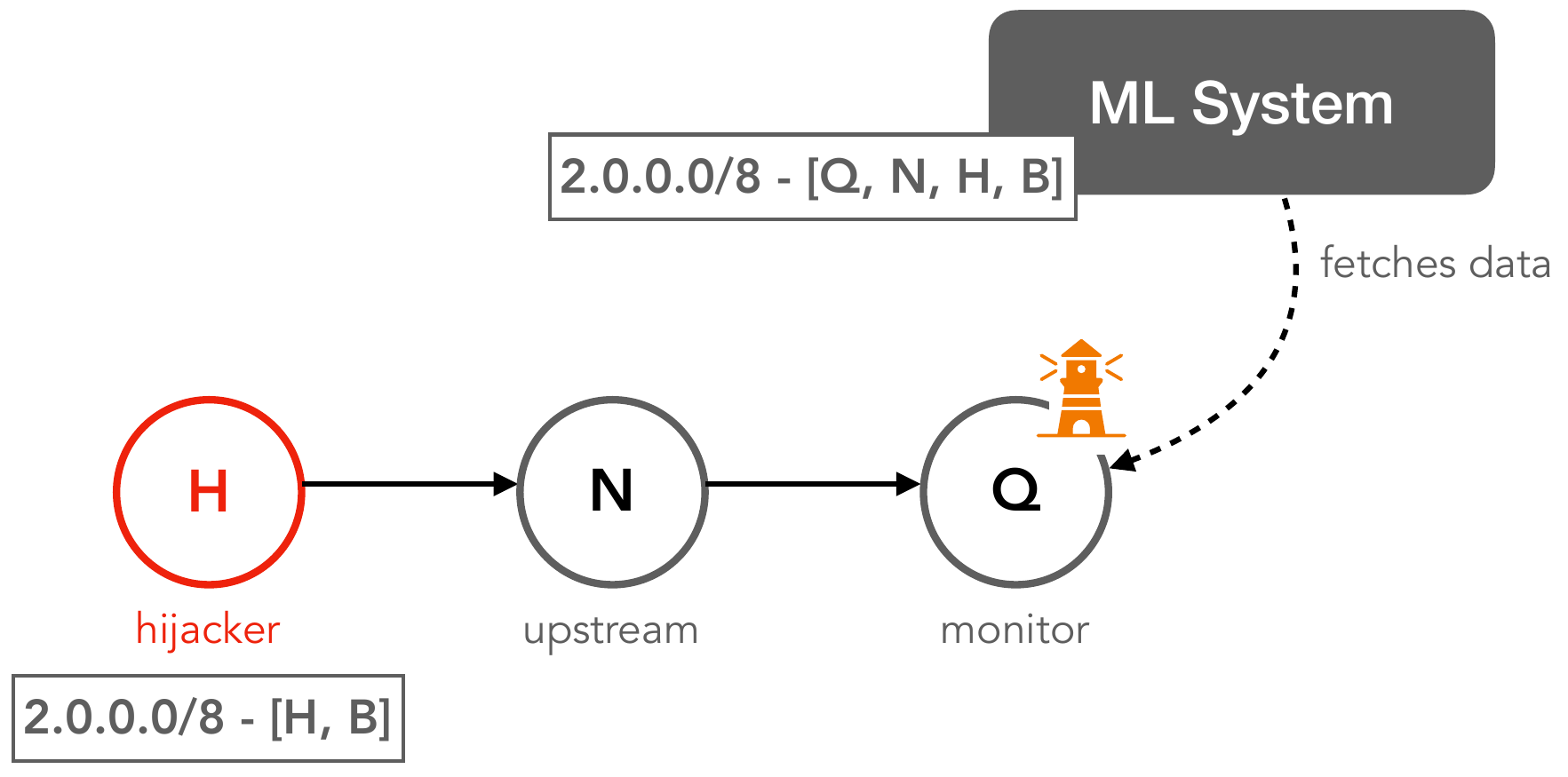}
    \caption{Monitor Poisoning Attack: Attacker H announces a prefix with forged origin B. The announcement propagates via H's upstream to a monitor. The ML system ingests this polluted data, corrupting its knowledge base.}
    \label{fig:monitor-pollution}
\end{figure}

Figure~\ref{fig:monitor-pollution} shows an example. The attacker in AS H announces a sub-prefix of a prefix owned by H, but falsely claims that AS B originated the route. B is not actually connected to H, but nobody can be sure of it -- except B, who will anyway discard the announcement according to the BGP's loop prevention mechanism. N cannot discard the announcement even if it is H's upstream provider, because it (i) cannot exclude that B is actually a new customer of H, and (ii) does not find an RPKI ROA for the announced sub-prefix. N thus accepts H's BGP route and propagates it further. Other ASes do the same as N. Eventually, the monitor in AS Q (as well as any other monitor) receives H's route and records the path containing the fake link H-B. When fetching data from monitors, monitor-based defenses ingest the fake route, and potentially add the fake link to their knowledge base.

Clearly, the next question is whether and how attackers can compute fake links that appear legitimate to hijack detectors, corrupt their knowledge bases and enable future hijacks.
We delve into this question in the following two sections.

\section{Attacking DFOH}
\label{sec:attack-dfoh}

\subsection{DFOH Architecture}
DFOH (Detecting Forged-Origin Hijacks) is designed to identify forged-origin BGP hijacks by analyzing AS paths observed in public BGP data \cite{dfoh}. Its architecture centers around a three-stage pipeline:

\begin{enumerate}
    \item \textbf{New Link Detection:} DFOH monitors BGP updates from public collectors (e.g., RouteViews, RIPE RIS) and compares observed AS links against a historical topology graph (built from $\approx$300 days of data). Links not present in the historical graph are flagged as ``new,'' triggering further analysis, as these often correlate with hijack events. The core idea is that forged-origin hijacks often introduce a new, previously unobserved link in the AS path, typically between the attacker and the (forged) origin AS.

    \item \textbf{Feature Computation:} For each new link, DFOH calculates a feature vector encompassing four categories to assess legitimacy:
    \begin{itemize}
        \item \textit{Topological:} Measures the impact of the new link on AS graph structure (e.g., centrality, neighborhood changes).
        \item \textit{Peering:} Infers peering likelihood based on shared infrastructure (IXPs, facilities) or geography, using data from sources like PeeringDB \cite{peeringdb} and focusing on neighbors' data to resist manipulation.
        \item \textit{AS-Path-Pattern:} Evaluates path validity against routing policy expectations (e.g., Gao-Rexford valley-free model \cite{gaorexford}) using learned models based on AS degree and customer cone sequences.
        \item \textit{Bidirectionality:} Checks if the link is observed in both directions (using BGP and IRR data), a strong indicator of legitimacy.
    \end{itemize}

    \item \textbf{Inference:} The computed features are fed into a Random Forest classifier trained to distinguish between legitimate links and forged-origin hijacks. This classifier is trained daily using a balanced sampling strategy that clusters ASes and ensures representative sampling across different AS types and potential attack scenarios, mitigating biases inherent in the AS topology. If multiple paths contain the new link, results are aggregated.
\end{enumerate}

The system aims to provide timely detection by focusing analysis on these newly observed links and leveraging a combination of topological, policy-based, and metadata features. However, as discussed in Section~\ref{subsec:fundamental-limitation}, its reliance on publicly observable data forms the basis for potential adversarial attacks.

\subsection{Poisoning DFOH's Knowledge Base}

While DFOH aims to detect forged-origin hijacks by identifying new AS links and analyzing associated features, its reliance on historical data gathered from public monitors creates a significant vulnerability - as attackers can manipulate the data ingested by these monitors. This allows for a knowledge base poisoning attack specifically tailored to circumvent DFOH's defenses.

The core principle of the attack is to strategically introduce carefully crafted, non-existent AS links into DFOH's historical topology graph \textit{before} launching an actual hijack. The goal is to manipulate the data against which future, malicious announcements are compared. The methodology involves the following steps:

\begin{enumerate}
    \item \textbf{Identify Poisonous Links:} The attacker identifies potential "poisonous" ASes. These are typically ASes not directly connected to the attacker but chosen strategically to manipulate DFOH's feature calculations favorably for a future hijack attempt against a target victim AS. The attacker leverages the small inherent error rate (false negatives) of the DFOH classifier, aiming for these crafted links to be misclassified as legitimate and added to the knowledge base.

    \item \textbf{Craft Poisoning Announcements:} The attacker announces a prefix they legitimately control (or a sub-prefix) but forges the origin AS in the BGP announcement to be one of the chosen poisonous ASes (see \ref{subsec:poisoning-monitors}).

    \item \textbf{Manipulate Feature Scores:} The poisonous links are chosen specifically to degrade DFOH's ability to detect a subsequent hijack involving the attacker (\textit{H}) and a victim (\textit{V}). For instance, by poisoning the knowledge base with links to ASes geographically close to the victim, the attacker can artificially lower the suspicion score derived from the PeeringDB features when the actual hijack link (H, V) appears later. As shown in Table~\ref{tab:feature_category_importance}, features like PeeringDB carry significant weight in the classification decision, making them effective targets for manipulation.

    \item \textbf{Execute Hijack:} After a sufficient period, allowing the poisoned data to be incorporated into DFOH's knowledge base (typically minutes), the attacker launches the actual forged-origin hijack. The presence of the previously injected poisonous links makes the new, malicious link appear less anomalous to DFOH's classifier, increasing the chance of evasion.
\end{enumerate}

By manipulating DFOH's input data, attackers can undermine its detection capabilities with minimal extra resources, posing a practical threat.

\begin{table}[ht]
    \centering
    \caption{Feature category importance scores for the DFOH Random Forest classifier, derived from training on data from 2024-03-01.}
    \label{tab:feature_category_importance}
    \begin{tabular}{lr}
    \hline
    \textbf{Feature Category} & \textbf{Importance Score} \\
    \hline
    ASPath Patterns & 0.59 \\
    PeeringDB & 0.23 \\
    Topological & 0.16 \\
    Bidirectionality & 0.02 \\
    \hline
    \end{tabular}
\end{table}

\subsection{Experimental Evaluation}
\label{ssec:dfoh-results}

\begin{figure*}
    \includegraphics[width=1.5\columnwidth]{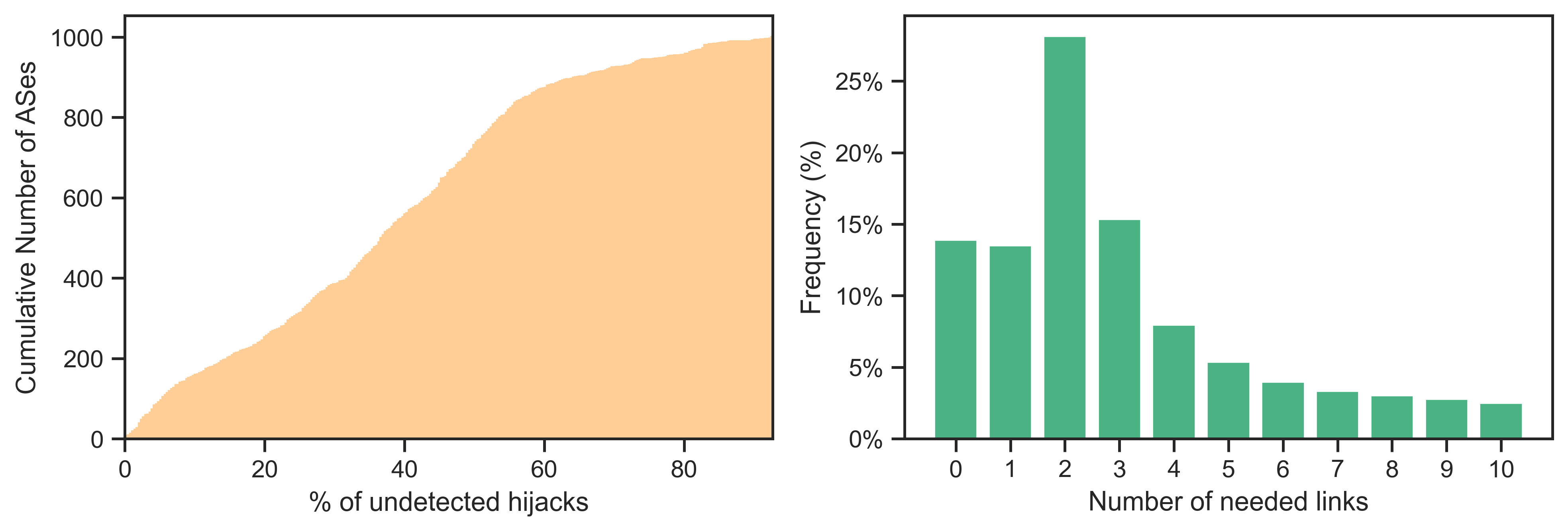}
    \centering
    \caption{Evaluation of the DFOH poisoning attack. Left: Distribution of hijack success rates across 1,000 simulated attackers targeting all other ASes. Each bin represents the percentage of successful hijacks achieved by one attacker after poisoning. Right: Distribution of the number of poisonous links required per attacker to achieve their respective success rates.}
    \label{fig:dfoh-data}
\end{figure*}

To evaluate the effectiveness of the knowledge base poisoning attack against DFOH, we conducted a large-scale simulation-based experiment. The simulations were performed on a cluster consisting of 2 nodes, each equipped with a double-socket Intel Xeon Gold 5118 CPU. While attacking a single target AS requires only a few seconds, the scale of the experiment - simulating 1,000 attacker ASes targeting each of the over 80,000 ASes in the internet topology - necessitated significant computational resources.

We simulated 1,000 attacker ASes, chosen randomly to represent a diverse set of network types and locations. For each simulated attacker, we attempted to poison DFOH's knowledge base to enable subsequent forged-origin hijacks against all other ASes. The simulation for each attacker utilized real-world AS paths observed from public collectors that originated from the attacker's AS. This represents a conservative approach, as many observed paths might be transient and not reappear, potentially underestimating the attack's effectiveness in a real-world scenario where attackers could sustain announcements.

The poisoning process involved identifying potential false negatives in DFOH's classification – links that DFOH would incorrectly classify as legitimate. From this set, we selected poisonous links using a heuristic designed to maximize the impact on DFOH's feature calculations for future hijack attempts. Specifically, we targeted features with high importance scores, such as PeeringDB (see Table~\ref{tab:feature_category_importance}), by selecting poisonous ASes geographically close to potential victim ASes. These crafted links were then assumed to be injected into the BGP data stream (as described in Section~\ref{subsec:poisoning-monitors}) and subsequently incorporated into DFOH's historical topology graph after being misclassified.

During initial simulations, we observed that some stub ASes, particularly those in remote or poorly connected regions, exhibited very low false negative rates according to DFOH's model. While seemingly positive for detection, this paradoxically implies that DFOH would likely generate frequent false positives for legitimate announcements originating from these networks. From the attacker's perspective, this finding initially limited the pool of viable poisonous AS candidates when targeting certain victims. To address this and model a more realistic attacker capability, we simulated the scenario where an attacker acquires additional connectivity through readily available commercial BGP transit services \cite{servperso, ifogch}. By announcing legitimate prefixes through a new provider connection, the attacker could expand their observed connectivity. Although DFOH's mechanism would initially flag and quarantine this new (legitimate) link for a period (30 days), an attacker willing to wait could subsequently leverage this expanded connectivity, significantly increasing their pool of potential poisonous ASes for manipulating the knowledge base. Our final evaluation incorporates this possibility.

Figure~\ref{fig:dfoh-data} presents the results of our simulation incorporating this enhanced attacker model. The left plot shows the distribution of hijack success rates achieved by each attacker after the poisoning phase. While the first 400 attackers achieved success rates below 40\%, a significant portion demonstrated high effectiveness. Notably, over 100 distinct attacker ASes were able to successfully hijack more than 80\% of all other ASes on the internet after poisoning DFOH's knowledge base.

The right plot in Figure~\ref{fig:dfoh-data} illustrates the number of poisonous links required to achieve these success rates. Crucially, the attack often requires only a small number of carefully chosen links. Approximately 15\% of the attackers needed zero additional links (which represents the false negatives), while around 14\% required only one link and 28\% needed two links. This demonstrates that attackers do not need to inject a large volume of malicious announcements to significantly compromise DFOH's detection capabilities.

These findings highlight the attack's practicality: by exploiting public data reliance, classification errors, and the potential to augment connectivity, a few crafted links can severely degrade DFOH's hijack detection.

\subsection{Naive Defense}
\label{ssec:dfoh-naive-defense}

A seemingly straightforward approach to counter the knowledge base poisoning attack might be to remove features from the DFOH model that are considered easily manipulable. For instance, given that PeeringDB information can be influenced by crafted announcements targeting specific geographic or infrastructural overlaps (as discussed in Section~\ref{subsec:poisoning-monitors}), one might consider removing the PeeringDB feature category entirely.

However, this naive defense strategy has significant drawbacks. Firstly, DFOH's strength lies in its combination of diverse feature categories, each contributing uniquely to detection accuracy across different scenarios, as indicated by their importance scores (Table~\ref{tab:feature_category_importance}). Removing a feature category, even a potentially vulnerable one, can significantly degrade the system's overall performance, leading to lower detection rates (True Positive Rate) and potentially higher false alarms (False Positive Rate) for legitimate events.

Secondly, attackers are not necessarily limited to manipulating only the most obvious features; they can craft announcements to influence topological features or AS-path patterns as well. A defense strategy based solely on removing features ignores the potential for broader manipulation and sacrifices detection capability. Therefore, simply removing potentially manipulable features is not a robust defense against strategic poisoning attacks.

\section{Attacking BEAM}
\label{sec:attack-beam}

\subsection{BEAM's Architecture}
\label{ssec:beam-arch}

BEAM (BGP sEmAntics aware network eMbedding) is a system designed to detect BGP anomalies by understanding the typical behavior of each AS within the Internet's routing landscape. Instead of just looking at path changes, BEAM learns these roles by analyzing the underlying structure of AS business relationships (like provider-customer or peer-to-peer links), which fundamentally dictate how routes are propagated \cite{beam}.

The core process involves two main steps:
\begin{enumerate}
    \item \textbf{AS Graph Construction:} BEAM utilizes datasets describing AS relationships (e.g., from CAIDA \cite{caida_rel}) to build a graph representing the inter-domain topology. In this graph, ASes are nodes, and directed edges represent relationships like provider-to-customer.

    \item \textbf{AS Embedding:} BEAM then employs a network representation learning model to generate a low-dimensional vector (an embedding) for each AS. This process is specifically designed to capture two key semantic properties derived from the AS graph:
    \begin{itemize}
        \item \textit{Proximity:} This measures how similar ASes are, considering both direct connections and shared neighbors (i.e., having similar connections to other ASes).
        \item \textit{Hierarchy:} This captures an AS's position within the Internet's structure, mainly based on provider-customer links. Tier-1 providers sit high in the hierarchy, while customer ASes are lower.
    \end{itemize}
    The model optimizes these embedding vectors so that the distance between vectors reflects the difference in the ASes' routing roles based on both proximity and hierarchy. The difference between any two AS roles can be measured using a function derived from these embeddings.
\end{enumerate}

BEAM uses these learned AS embeddings to evaluate BGP route changes. When a new route announcement for a prefix is observed, replacing a previous path, the system calculates a \textbf{path difference score}. This score quantifies the overall change in routing roles between the sequence of ASes in the old path and the new path. It is computed by comparing the embedding vectors of the ASes in both paths, often using an algorithm like Dynamic Time Warping (DTW) which measures the similarity between two temporal sequences \cite{dtw}.

To determine if a calculated path difference score indicates an anomaly, BEAM uses a \textbf{dynamic threshold}. This threshold is not fixed; it is periodically recalculated based on the distribution of path difference scores observed for route changes in a recent historical window (e.g., the previous hour). A path difference score exceeding this dynamic threshold is flagged as suspicious.

\subsection{Polluting the Threshold}
\label{ssec:beam-threshold-pollution}

Beyond manipulating the path characteristics themselves, BEAM's architecture presents a distinct vulnerability related to its dynamic threshold mechanism. Unlike the knowledge base poisoning attack effective against DFOH (Section~\ref{sec:attack-dfoh}), this attack targets the statistical basis used by BEAM to distinguish between normal and anomalous route changes.

The core idea of the threshold pollution attack is for an adversary to inject a controlled volume of crafted BGP announcements. These announcements are designed to generate path difference scores that fall just below the \textit{current} threshold, thereby initially evading detection. However, these injected "slightly abnormal but not anomalous" scores are then included in the pool of data used to calculate the threshold for the \textit{next} time window.

By persistently injecting such announcements, the attacker artificially inflates the statistics (e.g., mean, standard deviation) of the path difference scores considered "normal". This manipulation forces the system to calculate a higher threshold for the subsequent period. An elevated threshold makes it easier for the attacker's actual malicious announcements, such as those implementing a hijack (which inherently have significantly different routing roles and thus higher path difference scores), to fall below the inflated threshold and avoid being flagged as anomalous.

An attacker can anticipate or calculate the dynamic threshold because the inputs to its calculation are derived from publicly observable BGP data. By accessing public BGP monitoring feeds and potentially having access to a trained BEAM model or a functional equivalent, an attacker can observe the same route changes as the detection system and estimate the resulting threshold. This knowledge allows them to calibrate their injected announcements effectively for the pollution attack.

A potential challenge for this attack is injecting a sufficient volume of announcements within the observation window to significantly skew the statistics. However, the natural behavior of BGP provides an amplification factor. Due to routing oscillations and convergence processes, a single logical route announcement for a prefix is often repeated multiple times by routers across the network. Based on measurements from March 2024 \cite{routeviews2024}, a single prefix announcement was observed, on average, 6.43 times per hour (with a standard deviation of 17.79). This inherent repetition means an attacker only needs to initiate a relatively small number of distinct crafted announcements, as BGP dynamics will amplify their presence in the data streams monitored by systems like BEAM, making threshold manipulation feasible with less effort than might be initially assumed.

\subsection{Evaluation}
\label{ssec:beam-threshold-eval}

\begin{figure*}
    \centering
    \includegraphics[width=1.5\columnwidth]{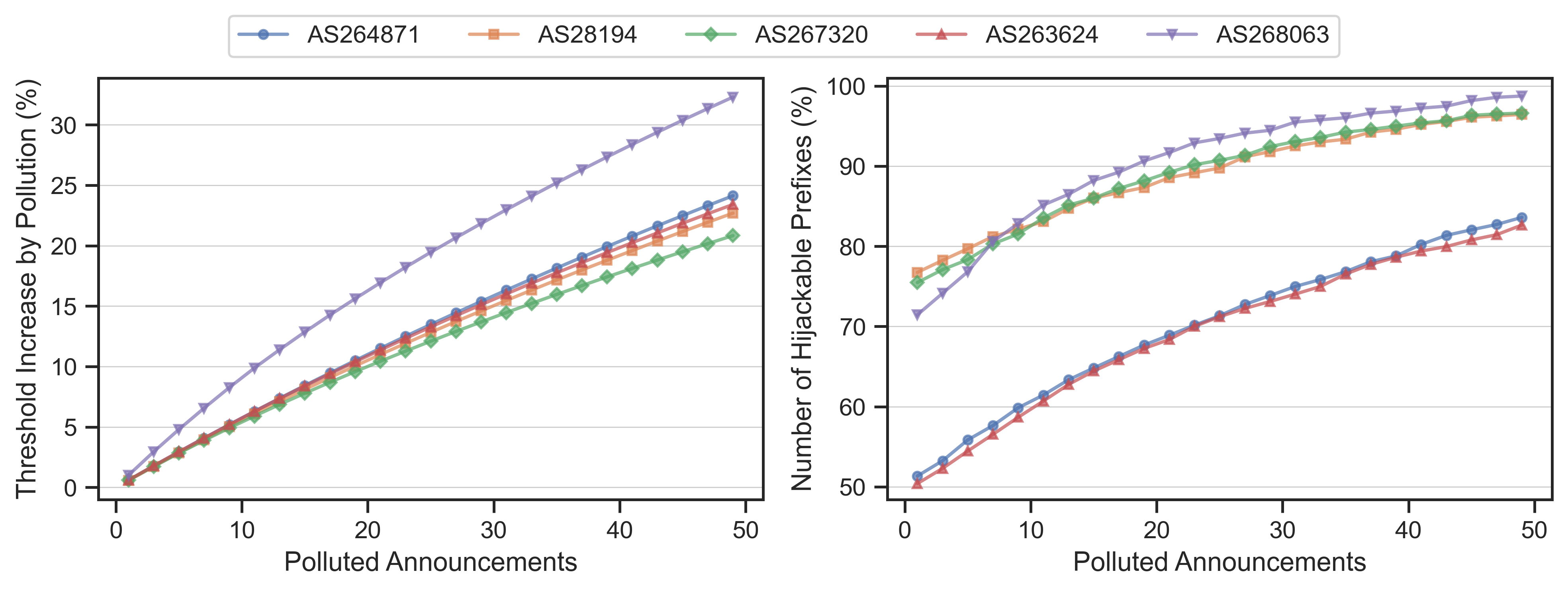}
    \caption{Effectiveness of the BEAM threshold pollution attack. The plot shows the percentage of prefixes an attacker can successfully hijack (evading detection) as a function of the number of distinct polluting announcements initiated.}
    \label{fig:beam-data}
\end{figure*}

We simulated the threshold pollution attack against BEAM to evaluate its practical impact. The simulation was performed on a machine equipped with an Apple M3 Pro CPU. We selected 5 random ASes to act as the attackers. For each attacker, we simulated the injection of a series of crafted BGP announcements. These announcements involved modifying legitimate paths originating from the attacker's AS by inserting a forged origin, carefully chosen such that the resulting path difference score would fall just below BEAM's current detection threshold. This strategy aims to have the announcements classified as legitimate while contributing to the statistics used for the next threshold calculation.

To model the amplification effect inherent in BGP, we analyzed real BGP update data from March 2024 \cite{routeviews2024} specifically for prefixes announced by the chosen attacker ASes. We measured the typical oscillation for these prefixes and used this observed rate to simulate how many times each distinct crafted announcement would likely appear in the monitoring data within the relevant time window. We varied the number of distinct polluting announcements initiated by the attacker from 1 to 50. After injecting the amplified set of announcements based on the observed oscillations, we recalculated BEAM's detection threshold based on the polluted data distribution. Finally, we determined the percentage of prefixes that could subsequently be hijacked using a forged-origin attack without triggering detection under the newly inflated threshold.

The results, illustrated in Figure~\ref{fig:beam-data}, demonstrate the effectiveness of this approach. We found that for most of the simulated attacker ASes, initiating just 10 distinct polluting announcements (amplified according to their observed prefix oscillation rates) was sufficient to raise the detection threshold by approximately 5\%. This seemingly modest increase in the threshold translated into a significant practical advantage for the attacker, allowing them to successfully hijack an additional 10\% of prefixes without being detected by BEAM compared to the baseline scenario without pollution. This highlights the practical risk posed by threshold pollution attacks, exploiting the statistical nature of the detection mechanism and the inherent amplification within BGP.

\subsection{Naive Defense}
\label{ssec:beam-naive-defense}

A seemingly straightforward defense against the threshold pollution attack described in Section~\ref{ssec:beam-threshold-pollution} would be to abandon the global dynamic threshold and instead maintain a separate, dynamically calculated threshold for each individual IP prefix. The intuition is that polluting the threshold for one prefix would not affect the thresholds for others, thus containing the attack's impact.

However, this per-prefix threshold approach, while conceptually simple, faces significant practical challenges that render it infeasible for a global monitoring system like BEAM:

\begin{itemize}
    \item \textbf{Scalability Issues:} The global routing table contains over one million IPv4 prefixes and over 220,000 IPv6 prefixes (as of mid 2025) \cite{cidrReport2025}. Maintaining, calculating, and storing a dynamic threshold for each of these prefixes would introduce immense computational and storage overhead, making the system difficult to scale and manage effectively.
    \item \textbf{Data Sparsity and Threshold Instability:} Calculating a reliable dynamic threshold requires a sufficient volume of historical data (legitimate route changes) for statistical significance. Many prefixes, especially more specific ones or those belonging to smaller organizations, exhibit very infrequent route changes. For such prefixes, there would be insufficient data within typical recalculation windows (like an hour) to establish a stable and accurate threshold. Attempting to calculate thresholds with sparse data would lead to high volatility, causing the threshold to fluctuate wildly and trigger constant false positives for minor, legitimate changes or, conversely, become too permissive and miss actual anomalies.
\end{itemize}

Therefore, while per-prefix thresholds might theoretically isolate pollution effects, the practical hurdles related to scalability and the need for robust statistical baselines based on sufficient data make this approach unworkable for a comprehensive BGP anomaly detection system.

\section{Countermeasures}
\label{sec:countermeasures}

The vulnerabilities exposed in monitor-based defenses necessitate exploring more robust security strategies. Long-term solutions aim to fundamentally enhance BGP's security architecture. Protocols like BGPSec \cite{bgpsec-rfc} offer cryptographic validation of the entire AS path, while alternative architectures like SCION \cite{scion-internetdraft} propose a complete overhaul of inter-domain routing with security built-in. However, despite significant research and efforts to incentivize adoption \cite{bgpsec-incentives-sig11, hlavacek2020disco}, widespread deployment of such fundamental changes faces immense practical hurdles due to the scale and inertia of the existing internet infrastructure. Modifying a protocol as pervasive as BGP remains a formidable challenge.

Given the slow progress of long-term solutions, short-term countermeasures focus on mitigating the risks within the current BGP ecosystem. One direction involves reducing the transparency of the detection system to potential attackers, essentially making it a "blackbox". By keeping the specific algorithms, features, thresholds, and potentially the training data confidential (e.g., in commercial offerings or non-open-source systems), defenders aim to make it significantly harder for attackers to probe the system for weaknesses, calculate false negative rates, or precisely tailor poisoning attacks like those demonstrated against DFOH and BEAM. However, this "security through obscurity" approach has inherent limitations. Determined attackers might still infer system behavior through careful observation or probing, and it hinders collaborative security research and independent verification.

Another, potentially complementary, short-term strategy involves augmenting public data with private BGP data feeds or out-of-band validation mechanisms. Systems like ARTEMIS \cite{artemis} exemplify this by combining publicly available monitor data with routing information directly obtained from the Routing Information Bases (RIBs) of participating routers. This allows for cross-validation; if a route change observed publicly is inconsistent with the private RIB data, it raises suspicion. However, even incorporating private data is not a panacea. While ARTEMIS and similar approaches can effectively detect anomalies for prefixes announced by ASes contributing private data, their visibility remains limited. Attackers can still exploit the vastness of the internet topology by using random or unused ASes, which are unlikely to be covered by private monitoring arrangements, to inject polluted routes or launch hijacks. This manipulation can still poison the public data components relied upon by hybrid systems or evade detection entirely if the attack path does not traverse the privately monitored infrastructure.

\begin{figure}[t!]
    \includegraphics[width=0.8\columnwidth]{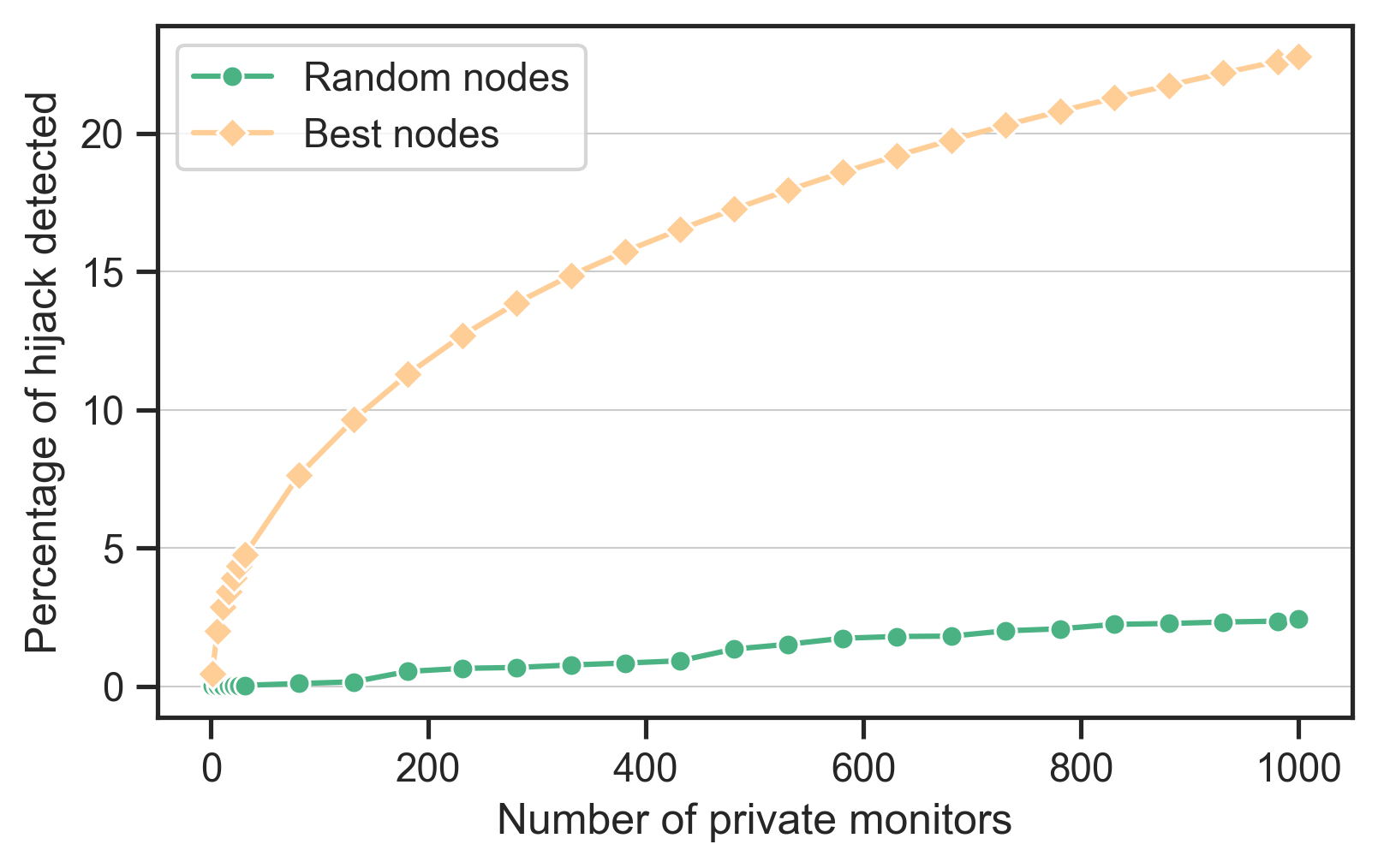}
    \caption{Hijack detection rate improvement when using an increasing number of private monitors, comparing random selection versus best-case selection.}
    \label{fig:defense-stats}
\end{figure}

To quantify the potential benefits and limitations of private monitoring, we simulated its impact on detecting the knowledge base poisoning attacks described earlier. The simulation was conducted on a machine equipped with an Apple M3 Pro CPU. We modeled scenarios where a varying number of ASes (from 1 to 1,000) act as private monitors, providing ground truth for routes they observe. The core logic involved checking if any of the poisoned links injected during the simulated DFOH attack (Section~\ref{ssec:dfoh-results}) would be directly observed by one of the designated private monitors. If a poisoned link involved a direct connection to a private monitor AS, it was considered detected. We compared two selection strategies: randomly choosing monitor ASes versus an optimal (best-case) selection that maximizes visibility into potential attack paths based on the simulated attack data. As shown in Figure~\ref{fig:defense-stats}, increasing the number of monitors improves detection. However, even with 1,000 randomly selected private monitors — a significant deployment — the detection rate for poisoned links remained below 3\%. This poor performance stems from the vast scale of the internet topology; with tens of thousands of ASes, the probability that an attacker's chosen AS for poisoning happens to have a direct peering relationship with one of the randomly placed private monitors is inherently low. The best-case selection strategy yielded better results, detecting over 20\% of attacks with 1,000 monitors, but still leaving a large fraction undetected. This highlights that even substantial private monitoring infrastructure struggles against attackers who can carefully choose where to inject malicious announcements, underscoring the challenge of achieving comprehensive BGP security through monitoring alone.

\section{Related Work}
\label{sec:relatedwork}

Securing the Border Gateway Protocol (BGP) has been a long-standing challenge. RPKI represents a significant effort to cryptographically validate the origin of route announcements \cite{rpkisurvey, rodday2023resource}. However, its incomplete deployment and inherent vulnerability to specific attacks like forged-origin hijacks necessitate complementary defense mechanisms. Consequently, BGP monitoring systems, often leveraging public data collectors like RouteViews \cite{routeviews} and RIPE RIS \cite{ripe-ris}, have become crucial. Early approaches focused on heuristic-based anomaly detection; simpler tools like BGPmon~\cite{bgpmon} and BGPalerter~\cite{bgpalert} alert operators to basic events like origin AS changes or RPKI-invalid announcements. Some systems, such as Cloudflare Radar~\cite{cloudflareradar}, incorporate additional data sources like Internet Routing Registries (IRR), though IRR data limitations regarding consistency and completeness restrict its reliability \cite{irrconsiderations, irrhygiene}. Other systems, like ARTEMIS \cite{artemis}, aim to improve detection speed and mitigation by combining public monitor data with private RIB feeds, addressing some limitations of purely public monitor-based approaches. More recent approaches increasingly incorporate machine learning (ML) to identify complex suspicious routing events. Systems like DFOH \cite{dfoh} utilize ML to detect forged-origin hijacks by analyzing path features derived from public data.

The application of ML in security domains, however, is fraught with challenges. Researchers have identified common pitfalls in the design, implementation, and evaluation of learning-based security systems, which can lead to unrealistic performance claims and hinder practical deployment \cite{pitfallsinml}. A major concern is the vulnerability of ML models to adversarial attacks, where malicious inputs are crafted to cause misclassification \cite{adversarialexamples, Biggio_2018}. Evaluating the robustness of ML models against such attacks is critical, especially in security contexts \cite{carlini2019evaluatingadversarialrobustness}. The threat of adversarial attacks against ML, specifically in network security,y has been surveyed, highlighting the adversarial nature inherent in tasks like intrusion and malware detection \cite{adversarialattacksnetworks}.

Within BGP security, research has explored how attackers can evade monitoring systems. Studies have demonstrated techniques to launch hijacks, such as sub-prefix hijacks using communities, that remain hidden from public monitors \cite{attackevadingmonitor}. Further investigations have analyzed the effectiveness of attackers deliberately crafting BGP paths to avoid propagation to route collectors, showing that even with expanded monitoring infrastructure, visibility gaps can persist \cite{evadingmonitors}. While these works address the fundamental limitations of monitor visibility, they do not specifically target the vulnerabilities within the ML models used by modern defense systems. Our work differs by focusing explicitly on the susceptibility of ML-based BGP hijack detection systems like DFOH and BEAM to adversarial manipulation, demonstrating how techniques like knowledge base poisoning and threshold pollution can undermine their effectiveness, an area that remains relatively underexplored.
\section{Conclusion}
\label{sec:conclusion}

This paper highlights a fundamental vulnerability of BGP monitors: the absence of route authentication makes the data they collect inherently unreliable. 
Worse still, sophisticated adversaries can actively manipulate this data to launch stealthier hijacks.
We demonstrate such manipulation against two recent BGP hijack detectors—DFOH and BEAM—showing that even minimal additional effort allows typical hijacks to evade detection.
These findings reveal a critical limitation of current defenses and emphasize the need to move beyond approaches that rely solely on public BGP data.

\newpage
\bibliographystyle{IEEEtran}
\bibliography{hotnets24-template}

\clearpage

\end{document}